\documentclass[12pt]{article}

\usepackage[superscript]{cite}
\usepackage{comment}
\usepackage[dvipsnames]{xcolor}
\usepackage{empheq}
\usepackage{amsmath}
\usepackage{bbold}
\usepackage{amssymb}
\usepackage{ulem,xpatch}
\usepackage{titling}
\usepackage{geometry} 
\usepackage{tabularx}

\newcommand{\cev}[1]{\reflectbox{\ensuremath{\vec{\reflectbox{\ensuremath{#1}}}}}}

\newcommand{\hide}[1]{\relax}

\newcommand{\Og}{\ensuremath{\Omega}}
\newcommand{\Om}{\ensuremath{\Omega_\mathrm{m}}}

\newcommand{\Gm}{\ensuremath{\Gamma_\mathrm{m}}}

\newcommand{\meff}{m_\mathrm{eff}}

\newcommand{\bnth}{\ensuremath{\bar n_\mathrm{th}}}

\newcommand{\bncav}{\ensuremath{\bar n_\mathrm{cav}}}

\newcommand{\bncavpr}{\ensuremath{\bar n_\mathrm{cav}}}

\newcommand{\Gmeas}{\ensuremath{\Gamma_\mathrm{meas}}}
\newcommand{\Gqba}{\ensuremath{\Gamma_\mathrm{qba}}}

\newcommand{\etameas}{\ensuremath{\eta_\mathrm{meas}}}

\newcommand{\xzpf}{\ensuremath{x_\mathrm{zpf}}}

\newcommand{\vcr}{\ensuremath{g_0}}

\newcommand{\etac}{\ensuremath{\eta_\mathrm{c}}}
\newcommand{\etadet}{\ensuremath{\eta_\mathrm{det}}}

\newcommand{\kpr}{\ensuremath{\kappa}}
\newcommand{\Dpr}{\ensuremath{\Delta}}

\newcommand{\Zpre}{\ensuremath{\vec{\bf r}}}

\newcommand{\nocontentsline}[3]{}
\newcommand{\tocless}[2]{\bgroup\let\addcontentsline=\nocontentsline#1{#2}\egroup}

\title{\bf \vspace{-2cm} Observing and Verifying \\ the Quantum Trajectory \\ of a Mechanical Resonator}
\author{\normalsize{Massimiliano Rossi$^{1, 2,\ast}$, David Mason$^{1, 2,\ast}$, Junxin Chen$^{1, 2}$, Albert Schliesser$^{1, 2, \dagger}$}\\
\small{\it $^{1}$Niels Bohr Institute, University of Copenhagen, 2100 Copenhagen, Denmark}\\
\small{\it$^{2}$Center for Hybrid Quantum Networks (Hy-Q), Niels Bohr Institute,}\\
\vspace{5mm}\small{\it University of Copenhagen, 2100 Copenhagen, Denmark}\\
\small{$^\ast$these authors contributed equally to this work}\\
\small{$^\dagger$to whom correspondence should be addressed; e-mail:  albert.schliesser@nbi.ku.dk}}
\date{}

\begin{document}
\maketitle
\paragraph*{Abstract\\}
Continuous weak measurement allows localizing open quantum systems in state space, and tracing out their quantum trajectory as they evolve in time.
Efficient quantum measurement schemes have previously enabled recording quantum trajectories of microwave photon and qubit states.
We apply these concepts to a macroscopic mechanical resonator, and follow the quantum trajectory of its motional state conditioned on a continuous optical measurement record. Starting with a thermal mixture, we eventually obtain coherent states of 78\% purity---comparable to a displaced thermal state of occupation 0.14.
We introduce a retrodictive measurement protocol to directly verify state purity along the trajectory, and furthermore observe state collapse and decoherence.
This opens the door to measurement-based creation of advanced quantum states, and potential tests of gravitational decoherence models.

\clearpage
\newpage
Within the Copenhagen interpretation of quantum mechanics, the quantum state of an isolated physical system is represented by its wavefunction.
This mathematical object encodes the probability of possible measurement outcomes and contains the maximum possible knowledge about the system.  
Under the usually inevitable coupling of the system to an unknown environment, the state evolves into a statistical mixture of quantum states, in a process known as decoherence.
The mixture is described via a density matrix $\rho$, which again encodes measurement probabilities, while accounting for the ignorance about system-environment interactions.
Decoherence entails the disappearance of some of the most salient, and useful, features of quantum mechanics, such as superposition and entanglement.
However, if information becomes available on how the system has interacted with the environment, it is possible to restore and retain the ‘purity’ of the quantum state (i.e. the extent to which the mixture is dominated by a single random, but known wavefunction).
Measurements can yield such information; over a finite time interval, however, the obtained information is often incomplete. \cite{Braginsky:1992aa, Wiseman:2010aa}
The density matrix can anyways be updated, conditioned on the particular measurement outcome, which purifies the state.
Sufficient measurement repetitions can then have the cumulative effect to project the system into a pure quantum state---akin to an ideal von Neumann measurement, which instantaneously ‘collapses’ the quantum state into a pure eigenstate of the measurement operator.
As the information accumulation through such a weak measurement takes time, obtaining pure conditional states require measurement rates that approach the system's total decoherence rate.
The latter may be notably increased by the presence of the measurement apparatus, which can itself be considered a bath which decoheres the system through its quantum backaction.

In the continuous limit of many weak subsequent measurements carried out over short times, the state conditioned on a measurement record traces out the system's {quantum trajectory} in time.
Observing pure quantum trajectories is a challenging task, yet achieved only in very clean settings such as cavity\cite{Guerlin:2007aa} and circuit\cite{Murch:2013aa,Weber:2014aa} QED.
Here, we extend these ideas to measurements of the motion of a macroscopic mechanical resonator\cite{Doherty:2012aa,Doherty:1999ab, Bowen:2016aa,Lammers:2018aa,Lammers:2018ab,Jacobs:2006aa}.
In this setting, pure conditional states are obtained through measurements of high efficiency $\eta_\mathrm{meas}=\Gmeas/\left(\gamma+\Gqba\right)$, where $\Gmeas$ is the measurement rate and $\gamma$ and $\Gqba$ are decoherence rates induced by a thermal bath, and measurement quantum backaction, respectively \cite{Aspelmeyer:2014aa, Clerk:2010ab}.
Prior experiments on motional state estimation has remained confined to the classical regime, due the fast decoherence by the thermal bath \cite{Vanner:2013aa, Wieczorek:2015aa, Harris:2016aa, Setter:2018aa}. 
In contrast, by probing the system with a measurement whose efficiency reaches $\eta_\mathrm{meas}\approx67\%$, we are able to observe individual quantum trajectories of highly-pure conditional states.  
Moreover, building on recent theoretical and experimental work on retrodiction \cite{Lammers:2018aa,Lammers:2018ab, Tan:2015aa, Zhang:2017aa} and past quantum states \cite{Gammelmark:2013aa, Rybarczyk:2015aa}, we introduce a retrodiction-based trajectory-verification technique, and use it to confirm the purity of the conditional state by statistical analysis of ensembles of trajectories.
This allows us to directly observe the collapse of the conditional state, as well as the decoherence that occurs in the absence of information from the measurement.

We explore these ideas in an optomechanical system based on an ultracoherent soft-clamped membrane resonator\cite{Tsaturyan:2017aa}.  The mechanical mode of interest (at the frequency $\Om/2\pi=1.14$~MHz) corresponds to a localized defect mode created within a phononic crystal.  This design simultaneously reduces radiative energy loss and avoids loss-inducing mode curvature, resulting in extremely low mechanical energy dissipation rate, $\Gm$.  For the device used here, we find $Q=\Om/\Gm=1.03\times10^9$ at temperature $T=$~11~K.
This motion is dispersively coupled to the frequency of a Fabry-Perot cavity mode (linewidth $\kappa/(2\pi)=$~18.5~MHz) at a characteristic vacuum optomechanical coupling rate $g_0/(2\pi)=$~129~Hz.  Populating the cavity with a large coherent field (with $\bncav$ average photons) leads to a linearized, field-enhanced coupling at rate $g=\sqrt{\bncav}g_0$. A probe laser which drives this cavity resonantly will acquire phase modulation sidebands proportional the mechanical displacement, which we detect via a balanced homodyne receiver.  The total detection efficiency (including cavity out-coupling) is $\etadet=$~74\%, ensuring that minimal mechanical information is lost.

In addition to a resonant probe beam, we also utilize an auxiliary beam to provide some pre-cooling of other modes of the membrane, via both sideband- and feedback-cooling\cite{Rossi:2018ab}. The effect of this beam on the main mode of interest is simply to change its effective thermal environment. Small, residual detuning of the probe similarly provides some damping.  In the following we account for both of these and refer to the effective energy damping rate and bath occupancy as, respectively, $\Gm/2\pi\approx$~130~Hz and $\bnth\approx$~2, such that $\gamma = \Gm\bar{n}_\mathrm{th}\approx2\pi\times260$~Hz.

The quantum measurement backaction of the probe is manifest as radiation pressure force fluctuations, leading to additional mechanical decoherence at rate $\Gqba=4g^2/\kappa\approx2\pi\times$2.54 kHz. Similarly, the mechanical displacement measurement can be characterized by a measurement rate $\Gmeas=4\etadet g^2/\kappa\approx2\pi\times$~1.88 kHz. Thus, the experimental system studied here can achieve measurements in which quantum backaction dominates thermal motion, and the measurement rate approaches the total decoherence rate (i.e. the measurement efficiency $\etameas=\Gmeas/(\Gqba+\gamma)\rightarrow1$).

\begin{figure}
\begin{center}
\includegraphics[scale=1]{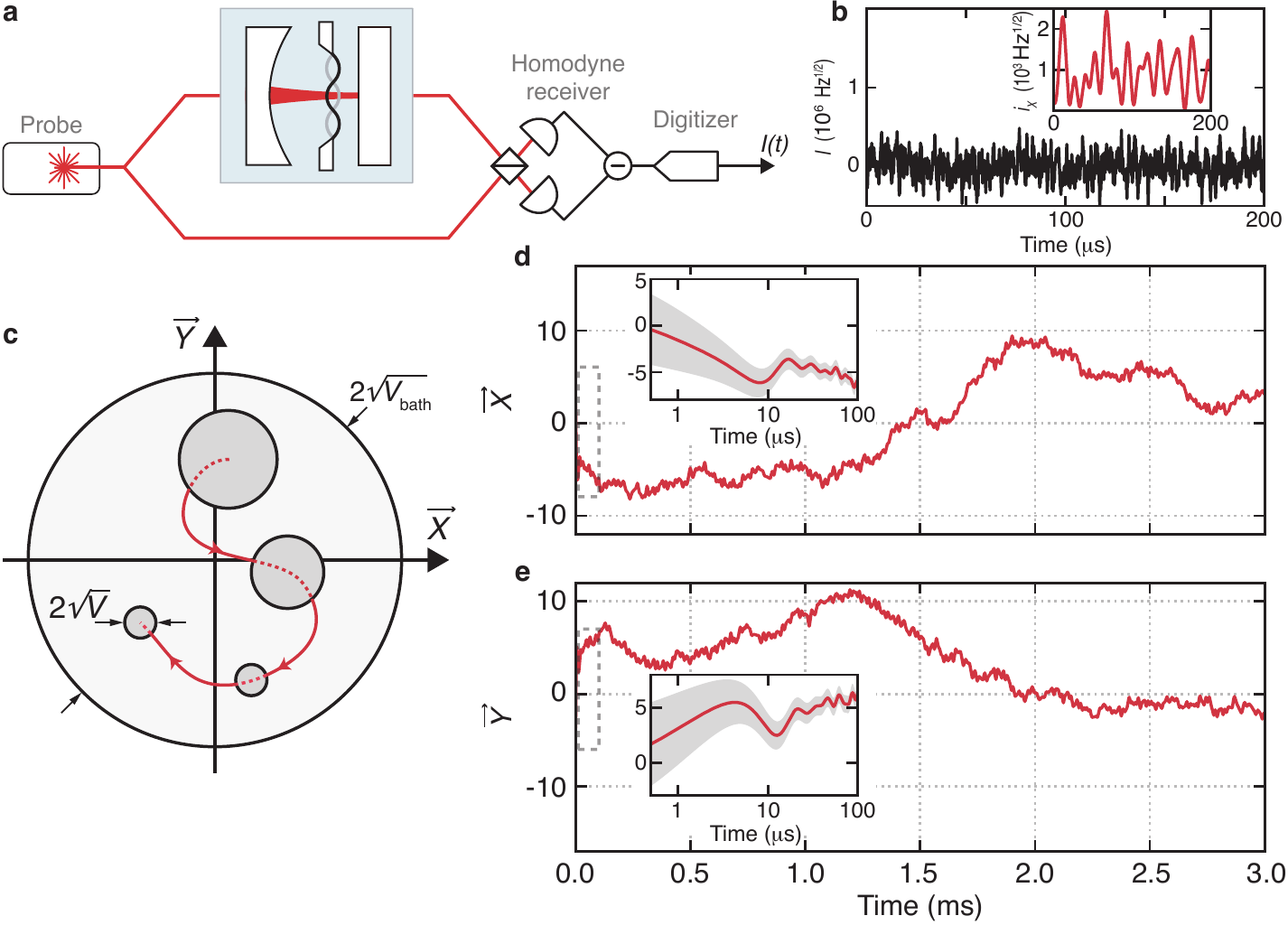}
\caption{{\bf Measuring a mechanical quantum trajectory.}
{\bf a,} Experimental setup. The mechanical resonator is coupled to an optical cavity, driven resonantly by a probe laser (red). The motion is imprinted on the phase quadrature of the transmitted light, which is measured with a balanced homodyne detector. The photocurrent $I(t)$ is digitized and analyzed in post-processing.
{\bf b,} Example calibrated photocurrent, containing information about all mechanical modes coupled to the cavity. Inset shows one quadrature signal obtained by demodulating the photocurrent at $\Om$.
{\bf c} Sketch of a quantum trajectory in phase space, in terms of the first moment $\vec{\mathbf{r}}$ (red line) and conditional variance (dark gray area, standard deviation). The variance is reduced as information is gathered during the measurements.
Averaging different realizations together leads to an unconditional, thermal state (light gray area), with variance $V_\mathrm{bath}$.
{\bf d, e} Measured single quantum trajectory $\vec{\mathbf{r}}(t)$, in terms of the slowly-varying quadratures, $\vec X(t)$ and $\vec Y(t)$. Insets illustrate the predicted decay of the conditional variance as the conditional state collapses (gray shaded area, standard deviation).
\label{f:fig_1}}
\end{center}
\end{figure}

The quantum trajectory of the mechanical resonator is derived from quadratures $\mathbf{i} = (i_X,i_Y)$ of the homodyne photocurrent, $I(t)$, demodulated at frequency $\Om$.  
This demodulation, which occurs in post-processing, uses a high-order low-pass filter whose $\sim$~120~kHz bandwidth is significantly larger than the total decoherence rate ($\gamma+\Gqba\approx2\pi\times$2.80~kHz), such that it has negligible effect on the mechanical signal\cite{Doherty:2012aa} (see Supplement).
The continuous data stream is subdivided into 3.2~ms segments, each of which is treated as an individual experimental realization. Examples of the raw photocurrent and one demodulated quadrature are shown in Fig.~\ref{f:fig_1}b.

These photocurrent quadratures form the measurement channels from which we extract the quantum trajectory.  This is done according to a stochastic master equation (SME), which simultaneuously describes unitary evolution of the system, environmental coupling, and Bayesian updates according to the measurement record\cite{Jacobs:2006aa, Belavkin:1980aa}.
Note that while our system is composed of both an optical and a mechanical mode, the cavity field can be adiabatically eliminated, since $\kappa\gg\Om$\cite{Doherty:2012aa, Hofer:2015ab}. For a high-Q resonator it is also convenient to move to the interaction picture at frequency $\Om$ and make a rotating wave approximation. Thus, we describe the system in terms of the slowly-varying quadratures $\hat{\mathbf{r}}=(\hat{X}$, $\hat{Y})$, where the mechanical position is $\hat{q}=\hat{X}\cos\left({\Om t}\right)+\hat{Y}\sin\left({\Om t}\right)$. The corresponding SME is \cite{Doherty:2012aa}
\begin{eqnarray}\label{e:sme}
d\rho &=& \left(\mathcal{L}_\mathrm{th} +\mathcal{L}_\mathrm{qba}\right)\rho dt + \sqrt{\Gmeas}\left(\mathcal{H}[\hat{X}]\rho\,dW_X+\mathcal{H}[\hat{Y}]\rho\,dW_Y\right) 
\end{eqnarray}
where $\mathcal{L}_\mathrm{th}$ and $\mathcal{L}_\mathrm{qba}$ describe interactions with the thermal and quantum optical baths, respectively (see Supplement). The final term, written in terms of the measurement superoperator $\mathcal{H}$ and two independent Wiener processes, $\mathbf{W}=(W_X, W_Y)$, describes the conditioning of the state.  This conditioning is based upon the measurement record, $\mathbf{i}\,dt = \sqrt{4\Gmeas}\vec{\mathbf{r}}dt + d\mathbf{W}$, where $\vec{\mathbf{r}} = \mathrm{tr}(\hat{\mathbf{r}} \rho)$ is a vector of the quadrature expectation values. Assuming Gaussian statistics, the conditional state is fully characterized by these expectation values along with the covariance matrix $V_{ij}=\mathrm{tr}(\{\hat{r}_i-\vec{r}_i,\hat{r}_j-\vec{r}_j \}\rho)/2$, where $\{\cdot\}$ is the anticommutator. For our purposes, this covariance can be written in terms of a single number, $V_{ij}=V\delta_{ij}$.

The dynamics of these first and second moments are given by
\begin{eqnarray}
d\Zpre &=&  -\frac{\Gm}{2}\Zpre dt + \sqrt{4\Gmeas}V(t) d\mathbf{W} ,  \label{e:Z_dyn}\\
\dot{V}(t) &=& - \Gm V(t) + \Gm (\bnth+\frac{1}{2}) +\Gqba - 4 \Gmeas V(t)^2 \label{e:V_dyn}
\end{eqnarray}

Note that while the expectation value evolution is stochastic (driven by the stochastic term $d\mathbf{W}$), the conditional variance evolves deterministically, decaying to a steady state value 

\begin{equation}
V = \frac{ \sqrt{1+16V_\mathrm{bath}\Gmeas/\Gm }-1}{8\Gmeas/\Gm},
\end{equation}

where $V_\mathrm{bath}=(\bnth+1/2 + \Gqba/\Gm)$ is the total bath variance\cite{Aspelmeyer:2014aa}. The reduction of the conditional variance (Fig. 2a) to its steady-state happens in a characteristic collapse time which depends inversely on the measurement rate\cite{Doherty:1999ab}. 
In the regime of fast, efficient measurement $\Gmeas\gg\gamma\gg\Gm$, relevant to this work, this variance approaches $V\approx1/(2\sqrt{\etameas})$. 
Thus, in the limit of highly efficienct measurements ($\etameas\rightarrow1$), the measurement process is able to project the initial thermal state into a pure coherent state ($V=1/2$).
We note that eqs.~\eqref{e:Z_dyn}-\eqref{e:V_dyn} are formally equivalent to a Kalman filtering problem\cite{Doherty:1999ab}, with constraints on the measurement and process noises imposed by quantum mechanics. Like the state estimate of a Kalman filter, the quantum state $\rho(t)$ enshrines the most accurate possible prediction of subsequent measurement outcomes.

In Fig.~\ref{f:fig_1}d, e we show an example of a quantum trajectory $\vec{\mathbf{r}}(t)$, as calculated according to eq.~\eqref{e:Z_dyn}. 
The region $0<t<100\,\mathrm{\mu s}$ indicates the collapse time of the measurement, after which the conditional variance has decayed to its steady state $V$. For the large probe strength used here (corresponding to $\etameas=$~67\%) the predicted conditional variance is $V=$~0.61, only 20\% larger than the zero-point fluctuations. This corresponds to a coherent state with purity $\mathcal{P}=\mathrm{tr}\left(\rho^2\right)=1/(2V)=$~0.82.\\

Unlike $\vec{r}$, this conditional variance $V$ cannot be immediately obtained from the experimental data: averaging individual trajectories approximates only the unconditional variance, $V_\mathrm{bath}$, in this regime (see Supplement and Fig.~\ref{f:fig_1}). To actually verify the prepared conditional state at time $t_0$, an experimenter could make a strong projective measurement at that time. 
Here we approximate this by a positive-operator valued measure (POVM) measurement based on subsequent data collected for $t>t_0$\cite{Gammelmark:2013aa, Zhang:2017aa}. To do so, we back-propagate an effect operator $E$ from future times to the past time $t_0$. The role of the effect operator is to refine, in a Bayesian sense, the probabilities for measurement outcomes, as determined by a density matrix $\rho(t_0)$. Together, $\rho$ and $E$ define the past quantum state\cite{Gammelmark:2013aa, Zhang:2017aa}, from which the expectation value of an operator $\hat{A}$ is calculated as $\mathrm{Tr}(\hat{A}\rho E)/\mathrm{Tr}(\rho E)$. The (unconditional) thermal state could be chosen for $\rho(t_0)$, disregarding the data collected before $t_0$.  If any prior (to $t_0$) information about the system is ignored, $\rho(t_0)\propto\mathbb{1}$, and $E$ contains all information about the quantum state at time $t_0$--as determined exclusively from measurements at later times.  We adopt this as a useful point of view, but note that the mathematical relations employed in the following comparisons hold irrespective of the interpretation of $E$.

For states whose evolution $\rho(t)$ is restricted to Gaussian states, the effect operator $E$ can be characterized by the expectation value $\cev{\mathbf{r}}=\mathrm{tr}(\mathbf{r} E)$ and covariance matrix $(V_E)_{ij}=\mathrm{tr}(\{\hat{r}_i-\cev{r}_i,\hat{r}_j-\cev{r}_j \}E)/2$.
For the mechanical measurement performed here,  $\mathbf{V_E}$ can be written as $(V_E)_{ij}=V_E\delta_{ij}$, and the first and second moments evolve according to\cite{Lammers:2018aa, Lammers:2018ab,Zhang:2017aa}
\begin{eqnarray}
-d\,\cev{\mathbf{r}} &\equiv& \cev{\mathbf{r}}(t-dt)-\cev{\mathbf{r}}(t) =  \frac{\Gm}{2}\cev{\mathbf{r}} dt + \sqrt{4\Gmeas}V_E(t) d\mathbf{W}_E, \label{e:r_E}\\
-\dot{V_E}(t) &\equiv& \frac{V_E(t-dt)-V_E(t)}{dt} = \Gm V_E(t) + \Gm (\bnth+\frac{1}{2}) +\Gqba - 4 \Gmeas V_E(t)^2,
\label{e:V_E}
\end{eqnarray}
where $d\mathbf{W}_E = \mathbf{i} dt - \sqrt{4\Gmeas}\,\cev{\mathbf{r}} dt$ is a stochastic variable and the steady state conditional variance is $V_E=V+\Gm/(4\Gmeas)\approx V$.
It is this retrodicted trajectory (determined only by the measurement record after $t_0$) which we will use to verify individual quantum trajectories such as the one shown in Fig.~\ref{f:fig_1}. 

\begin{figure}
\begin{center}
\includegraphics[scale=1]{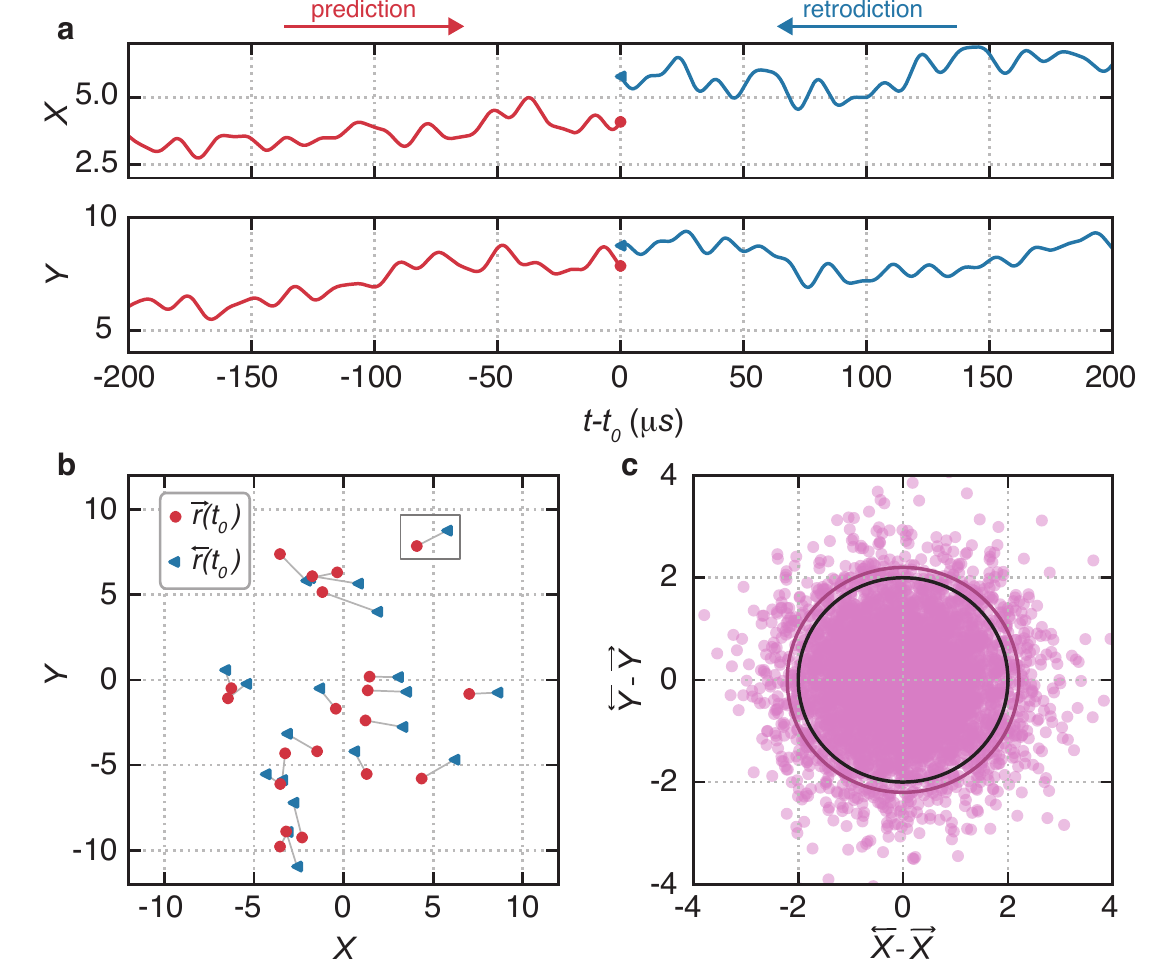}
\caption{{\bf Verification of the conditional state.}
{\bf a} A single quantum trajectory, $\protect\vec{\mathbf{r}}(t)$ (red line), calculated until time $t_0$, is compared with a retrodiction, $\protect\cev{\mathbf{r}}(t)$ (blue line), back-propagated to the same time $t_0$.
{\bf b} An ensemble of relative state estimations at time $t_0$, as measured from the trajectory $\vec{\mathbf{r}}(t_0)$ (red circles) and verified by the retrodiction $\protect\cev{\mathbf{r}}(t_0)$ (blue triangles). The prediction-retrodiction pairs are connected by a gray line. The gray box indicates the pair shown in {\bf a}.
{\bf c} Phase space distribution of the relative expectation values, $\protect\cev{\mathbf{r}}(t_0)-\vec{\mathbf{r}}(t_0)$ (purple circles). The purple line corresponds to two standard deviations of the data (radius 2$\sqrt{\sigma^2}$), compared with the expected $\sqrt{\sigma^2}$ for a pure coherent state (black line, radius $2\sqrt{1}$).
\label{f:fig_3}}
\end{center}
\end{figure}

In Fig.~\ref{f:fig_3}a we show a predicted and retrodicted trajectory, $\vec{\mathbf{r}}(t)$ and $\cev{\mathbf{r}}(t)$, in a time interval in which both conditional variances have reached the steady-state.
These trajectories are compared at an arbitrary common end point, $t_0$.
Various pairs of $\vec{\mathbf{r}}(t_0)$ and $\cev{\mathbf{r}}(t_0)$, from different experimental trials, are shown in Fig.~\ref{f:fig_3}.
We calculate, over this ensemble of experimental realizations, the covariance matrix of the relative trajectories $\pmb{\sigma}^2 = \mathrm{Cov}\left[\vec{\mathbf{r}}(t_0)-\cev{\mathbf{r}}(t_0)\right]$. 
 
As expected, we always find that $\sigma^2_{XY}\approx0$ and $\sigma^2_{XX}\approx\sigma^2_{YY}$ (within 2\%), and henceforth report only the average diagonal term, $\sigma^2$.
This \textit{experimental} variance provides a direct measurement of the desired \textit{quantum} conditional variance $V$. As we derive in the Supplement, the ensemble variance is given by a simple sum of the variances of the operators characterizing the pre- and retrodicted quantum state, respectively,
\begin{equation}
\sigma^2=V+V_E\approx2V
\end{equation}
a result quite compatible with intuition.
We experimentally find a variance of $\sigma^2 = 1.29$, which agrees with the predicted $\sigma^2=V+V_E=1.24$  to within 4\%, and corresponds to a purity $\mathcal{P} = 0.78$.
(This includes a correction due to 6\% systematic error introduced by demodulation filter correlations -- see Supplement).
This corresponds to a displaced thermal state of occupation $\bar{n}_\mathrm{cond}=0.14$.  In this sense, this process is sometimes referred to as ``cooling by measurement''\cite{Vanner:2013aa}.
Force feedback based on the predicted quantum state can in principle entirely undo the displacement, to yield a zero-mean, low-entropy state\cite{Doherty:1999aa, Doherty:2012aa}.

\begin{figure}
\begin{center}
\includegraphics[scale=1]{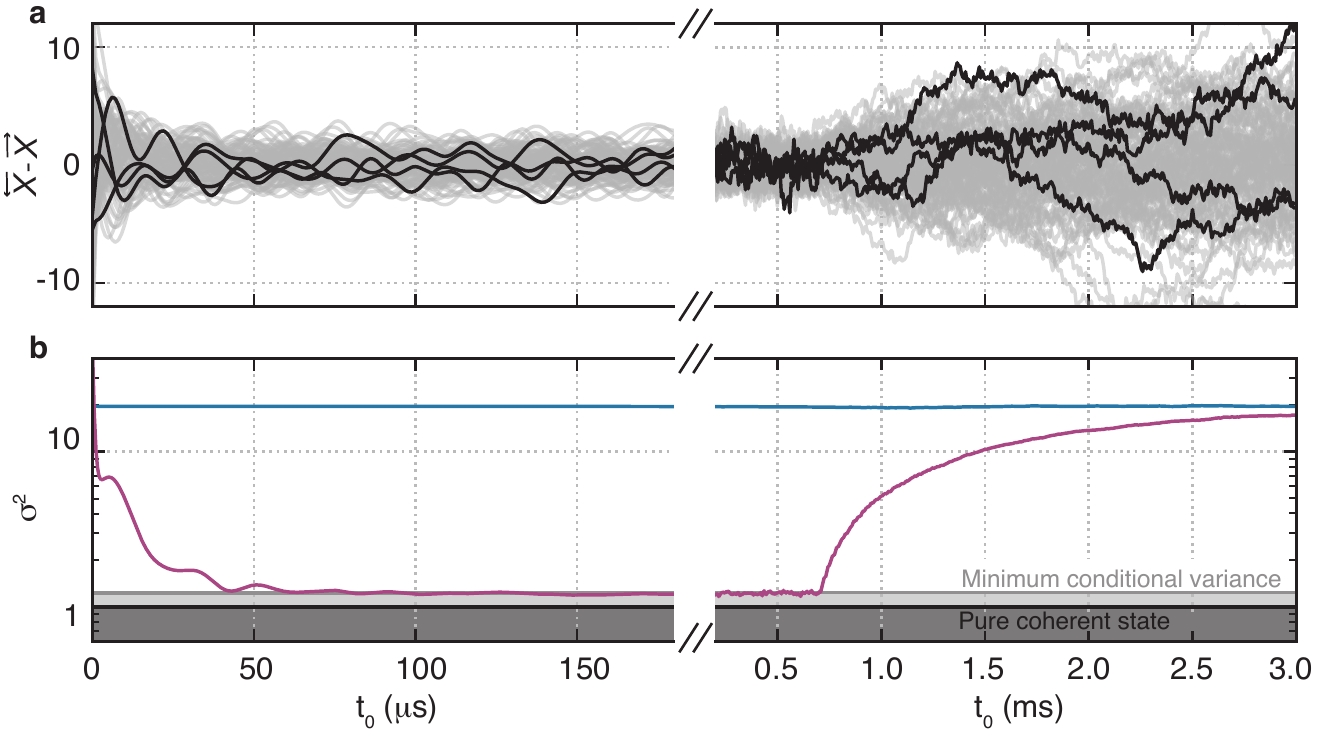}
\caption{{\bf Measurement-induced collapse and decoherence.}
{\bf a} $X$-quadratures of relative trajectories $\protect\cev{\mathbf{r}}(t)$ - $\protect\vec{\mathbf{r}}(t)$ (gray lines), shown during different time intervals. A few traces are highlighted in black for illustration purposes.
{\bf b} The relative variance ($\sigma^2$, purple) of the ensemble of relative trajectories from {\bf a}. For comparison, the unconditional variance (calculated from $\protect\cev{\mathbf{r}}(t)$ only) is shown in blue. The theoretical expectation ($V+V_E$) is shown in gray and the variance of a pure coherent state is shown in black.
\label{f:fig_4}}
\end{center}
\end{figure}

We can extend this retrodiction-verification protocol to study the measurement process in more detail, including its dynamics.  
In particular, we can observe both the measurement-induced collapse of the conditional state, and decoherence in the absence of measurement conditioning.  To do so, we compare retrodicted quadrature values with the forward-calculated ones at a time $t_0$, which now varies within the interval $0<t_0<3$~ms.  
The resulting relative trajectories ($\vec{\mathbf{r}}(t_0)-\cev{\mathbf{r}}(t_0)$) and their ensemble variance are shown in Figs.~\ref{f:fig_4}a, b, respectively. 
Note that the retrodiction always begins at 3.2~ms, such that its conditional variance is in the steady-state throughout the displayed time interval.  
In contrast, the predictions all begin at $t=0~\mathrm{\mu s}$. Thus, up to $\approx50~\mathrm{\mu s}$, the conditional variance $V(t)$ is expected, per eq.~\eqref{e:V_dyn}, to decay from an unconditional thermal state. Indeed, this is exactly the behavior revealed by the retrodictive state verification in Fig.~\ref{f:fig_4}b. 
Next, to visualize the dynamics of the system in the absence of measurement, we set the measurement efficiency to zero in eq.~\eqref{e:Z_dyn} from time 0.7~ms onward. That is, we stop conditioning the predicted quantum trajectory based on the measurement record.  Thus, the conditional variance rethermalizes, and our nearly-pure state decoheres into a statistical mixture.\\

The ability to effect strong projective displacement measurements via continuous weak measurement opens the door to various measurement-based protocols. Applying the same underlying machinery to modified measurement schemes allows, for example, the production of conditional mechanical squeezed states, as well as conditional entanglement and quantum steering\cite{Danilishin:2013aa, Lammers:2018aa, Lammers:2018ab}. Moreover, while the analysis presented here was implemented in post-processing of data, there is no fundamental obstacle to implementing this protocol as a real-time filter (e.g. using an FPGA). Real-time feedback that removes the (known) mean displacement can then yield unconditional states with the desired quantum correlations. Finally, the measurement and analysis techniques presented here could be of interest to test and constrain models for gravitationally induced decoherence, such as continuous spontaneous localization\cite{Nimmrichter:2014aa, Bassi:2017aa, McMillen:2017aa}.

\paragraph*{Acknowledgements}
The authors would like to acknowledge helpful discussions with Klemens Hammerer, Klaus M{\o}lmer and Marco G. Genoni, and sample fabrication by Yeghishe Tsaturyan. This work was supported by funding from the European Union’s Horizon 2020 research and innovation programme (European Research Council (ERC) project Q-CEOM, grant agreement no. 638765 and FET proactive project HOT, grant agreement no. 732894), a starting grant from the Danish Council for Independent Research and the Carlsberg Foundation.

\newpage
\renewcommand{\figurename}{{\bf Supplementary Fig.}}
\renewcommand{\tablename}{{\bf Supplementary Table}}
\setcounter{figure}{0}\renewcommand{\thefigure}{{\bf S\arabic{figure}}}
\setcounter{table}{0}\renewcommand{\thetable}{{\bf S\arabic{table}}}
\setcounter{equation}{0}\renewcommand{\theequation}{S\arabic{equation}}
\paragraph{{\LARGE Supplementary Information}}

\paragraph*{Stochastic Master Equation}
In the stochastic master equation, eq.~\eqref{e:sme}, the state evolution is influenced by thermal and optical baths, represented by superoperators, 
\begin{align}
    \mathcal{L}_\mathrm{th}\rho &= \Gm\bar{n}\mathcal{D}[\hat{c}^\dagger]\rho + \Gm(\bar{n}+1)\mathcal{D}[\hat{c}]\rho\,\,\text{and}\\
    \mathcal{L}_\mathrm{qba}\rho &=\Gqba\left(\mathcal{D}[\hat{c}^\dagger]\rho+\mathcal{D}[\hat{c}]\rho\right),
\end{align}
respectively. The latter is also referred to as the quantum backaction of the measurement. These superoperators act on the slow mechanical amplitude $\hat{c} =\left(\hat{X} + \imath\hat{Y}\right)/\sqrt{2}$. 
The dissipation and measurement superoperators are, respectively,
\begin{align}
  \mathcal{D}[\hat{a}]\rho &= a\rho \hat{a}^\dagger - (\hat{a}^\dagger \hat{a}\rho + \rho \hat{a}^\dagger \hat{a})/2\,\, \text{ and} \\
  \mathcal{H}[\hat{a}]\rho &= \hat{a}\rho + \rho \hat{a}^\dagger - \mathrm{tr}\left((\hat{a}+\hat{a}^\dagger) \rho\right) \rho.
\end{align}

\paragraph*{Connection between experimental averages and quantum expectation values}
We want to connect statistical quantities which we can calculate from an ensemble of experiments to quantum expectation values and variances. 
To avoid confusion we define the quantum expectation value (QEV) of an operator $\hat{A}$ as $\vec{A}=\mathrm{tr}\left(\rho\hat{A}\right)$ if based on a quantum state $\rho$ or $\cev{A}=\mathrm{tr}\left(E\hat{A}\right)$ if based on the effect operator, $E$.
Hence, in our notation, arrows indicate pre- and retrodicted first moments of an operator, while we generally denote vector quantities in \textbf{boldface}.
Instead, the classical expectation value (CEV), or average, of a stochastic variable $r$ is defined as $\left<r\right>$ and its variance $\mathrm{Var}[r]\equiv\left<\left(r-\left<r\right>\right)^2\right>$.

Let's call $\mathbf{\hat{r}}=(\hat{X}, \hat{Y})$ the vector of the mechanical slow quadrature operators.  The stochastic master equation then predicts the evolution
\begin{equation}
    d\vec{\mathbf{r}} = -(\Gm/2)\vec{\mathbf{r}} dt +  \sqrt{4\Gmeas} V(t) d\mathbf{W}
\end{equation}
for the first moments, i.e., the quantum expectation values $\vec{\mathbf{r}}=\mathrm{tr}(\rho\, \mathbf{\hat{r}})$.
Here, $\mathbf{W}$ is a Wiener process, i.~e.~a stochastic (vector) variable satisfying $\left<dW_i\right>=0$ and $\left<dW_idW_j\right>=\delta_{ij}dt$, and $V(t)$ is the time-dependent conditional variance of the quantum state. This variance satisfies a Riccati equation and for an initial thermal state with variance $V_\mathrm{bath} = \bnth + 1/2 + \Gqba/\Gm$ its solution is 
\begin{equation}
V(t) = V + \frac{2V + \Gm/(4\Gmeas)}{e^{(8V\Gmeas + \Gm)t}\left[1+\Gm/(4\Gmeas V)\right]^2-1},
\end{equation}
where $V$ is the steady state conditional variance given by\cite{Bowen:2016aa, Doherty:2012aa}
\begin{equation}
    V=\frac{\sqrt{1+16V_\mathrm{bath}\Gmeas/\Gm  }-1}{8 \Gmeas/\Gm}.
\end{equation}
We henceforth assume that such steady state regime has been reached (i.e. the observation time is much longer than the collapse time). We then formally solve the first moment equations and find
\begin{eqnarray}\label{e:rt_sol}
\vec{\mathbf{r}}(t) = e^{-\frac{\Gm}{2}\left(t-t_i\right)} \vec{\mathbf{r}}(t_i) + \sqrt{4 \Gmeas}V \int_{t_i}^t e^{-\frac{\Gm}{2}\left(t-t'\right)} d\mathbf{W}(t').
\end{eqnarray}
These first moments $\vec{\mathbf{r}}(t)$ are directly obtained from the measurement record and represent a stochastic, classical variable and we can thus calculate statistics in terms of CEVs. From eq.~\eqref{e:rt_sol} and $\left<d\mathbf{W}\right>=0$ we have $\left<\vec{\mathbf{r}}(t)\right> = e^{-(\Gm/2)(t-t_i)}\left<\vec{\mathbf{r}}(t_i)\right>$. 
The assumption of the conditional state variance to be in the steady state corresponds to assuming the initial condition to be far in the past, i.e. $t_i\rightarrow-\infty$. In this limit, we can also disregard the first term in the right hand side of eq.~\eqref{e:rt_sol} and find the expected result $\left<\vec{\mathbf{r}}(t)\right> = 0$.

Let's calculate now the second-order correlation function of the random variable $\vec{\mathbf{r}}(t)$,
\begin{eqnarray}
\left<\vec{\mathbf{r}}(t)\vec{\mathbf{r}}(t')\right> = 4\Gmeas V^2 \left< \int_{-\infty}^t e^{-\frac{\Gm}{2}\left(t-\tau\right)}d\mathbf{W}(\tau)\int_{-\infty}^{t'} e^{-\frac{\Gm}{2}\left(t'-\tau'\right)} d\mathbf{W}(\tau')\right>.
\end{eqnarray}
Exploiting It\^o isometry, for $t'>t$ we find
\begin{eqnarray}\label{e:rt_rtp}
\left<\vec{\mathbf{r}}(t)\vec{\mathbf{r}}(t')\right> = 4\Gmeas V^2 e^{-\frac{\Gm}{2}\left(t+t'\right)} \int_{-\infty}^t e^{\Gm \tau}d\tau = 4\frac{\Gmeas}{\Gm}V^2 e^{-\frac{\Gm}{2}\left|t'-t \right|}.
\end{eqnarray}
At equal time we find 
\begin{eqnarray}
\left<\vec{\mathbf{r}}(t)^2\right> =  4\frac{\Gmeas}{\Gm} V^2.
\end{eqnarray}

In the limit of fast, efficient measurement $\Gmeas\gg\gamma\gg\Gm$, an average over experimental realizations then recovers the expected unconditional, steady-state quantum variance determined by thermal noise and quantum backaction heating, $\left<\vec{\mathbf{r}}(t)^2\right> \approx \bar n_\mathrm{th}+1/2+\Gqba/\Gm$.

To extract the \emph{conditional} variance, we compare with retrodictions of the quantum state at the same time $t$. The QEV of $\hat{\mathbf{r}}$ from the retrodicted effect operator evolves according to\cite{Zhang:2017aa, Lammers:2018aa, Lammers:2018ab}
\begin{equation}
  -d\,\cev{\mathbf{r}} \equiv\cev{\mathbf{r}}(t-dt)-\cev{\mathbf{r}}(t)=( \Gm/2)\,\cev{\mathbf{r}} dt + \sqrt{4\Gmeas} V_E(t) d\mathbf{W}_E,
\end{equation}
where $d\mathbf{W}_E(t)=\mathbf{i}(t) dt - \sqrt{4\Gmeas} \cev{\mathbf{r}} dt$. 
We assume that the final condition is far in the future, so the quantum variance $V_E(t)$ is in the steady state and given by $V_E =V + \Gm/\left(4\Gmeas\right)$. 
To express the random variable \cev{\mathbf{r}} in terms of the Wiener stochastic process, we use $d\mathbf{W}_E(t)=\mathbf{i}(t) dt - \sqrt{4\Gmeas} \cev{\mathbf{r}} dt = \sqrt{4\Gmeas}(\vec{\mathbf{r}}-\cev{\mathbf{r}})dt + d\mathbf{W}(t)$. The formal solution is 
\begin{eqnarray}\label{e:rt_E_sol}
\cev{\mathbf{r}}(t) =  4\Gmeas V_E\int_t^{\infty}e^{-\lambda\left(\tau-t\right)}\vec{\mathbf{r}}d\tau + \sqrt{4\Gmeas}V_E \int_t^{\infty}e^{-\lambda\left(\tau-t\right)} d\mathbf{W}(\tau),
\end{eqnarray}
where $\lambda=4\Gmeas V_E - \Gm/2$. The first moment is $\left<\cev{\mathbf{r}}(t)\right> = 0$. The variance of this process is
\begin{eqnarray}
\left<\cev{\mathbf{r}}(t)^2\right> &=& (4\Gmeas V_E)^2 e^{2\lambda t} \left<\int_t^\infty e^{-\lambda \tau} \vec{\mathbf{r}}(\tau)d\tau\int_t^\infty e^{-\lambda \tau'} \vec{\mathbf{r}}(\tau')d\tau'\right> + \\ \nonumber
&+&4\Gmeas V_E^2 e^{2\lambda t} \left<\int_t^\infty e^{-\lambda \tau} d\mathbf{W}(\tau)\int_t^\infty e^{-\lambda \tau'} d\mathbf{W}(\tau')\right> + \\ \nonumber
&+&2\times4\Gmeas\sqrt{4\Gmeas}V_E^2 e^{2\lambda t} \left<\int_t^\infty e^{-\lambda \tau}\vec{\mathbf{r}}(\tau)d\tau \int_t^\infty e^{-\lambda \tau'} d\mathbf{W}(\tau')\right>,
\end{eqnarray}
where we have broken down the three contributions. The first integral (using eq.~\eqref{e:rt_rtp}) gives
\begin{eqnarray}
(I) = \frac{8\Gmeas V_E}{8\Gmeas V_E-\Gm} \,4\frac{\Gmeas}{\Gm}V^2,
\end{eqnarray}
the second integral, using It\^{o} isometry, is
\begin{eqnarray}
(II) = \frac{8\Gmeas V_E}{8\Gmeas V_E-\Gm} \frac{V_E}{2}
\end{eqnarray}
and finally the third integral, using eq.~\eqref{e:rt_sol} and the It\^{o} isometry again, is
\begin{eqnarray}
(III) = \frac{8\Gmeas V_E}{8\Gmeas V_E-\Gm} V.
\end{eqnarray}
Combining these terms together and using the fact that $V=V_E - \Gm/(4\Gmeas)$, we find that
\begin{eqnarray}
  \langle\cev{\mathbf{r}}(t)^2\rangle = 4 \frac{\Gmeas}{\Gm} V_E^2
\end{eqnarray}

Finally, for our retrodiction-verification protocol, we are interested in the cross-correlation between $\vec{\mathbf{r}}$ and $\cev{\mathbf{r}}$, i.e.
\begin{eqnarray}
\left<\vec{\mathbf{r}}(t)\,\cev{\mathbf{r}}(t)\right> = 4\Gmeas V_E e^{\lambda t}\left< \int_t^\infty e^{-\lambda t'}\vec{\mathbf{r}}(t)\vec{\mathbf{r}}(t') dt'\right> = 4 \frac{\Gmeas}{\Gm} V^2.
\end{eqnarray}
We have now all the tools to calculate $\sigma^2(t) = \mathrm{Var}\left[\cev{\mathbf{r}}(t)-\vec{\mathbf{r}}(t)\right]$. Using the fact that $V_E=V+\Gm/(4\Gmeas)$, we find
\begin{eqnarray}
\sigma^2(t) = \mathrm{Var}\left[\cev{\mathbf{r}}(t)\right] + \mathrm{Var}\left[\vec{\mathbf{r}}(t)\right] -2\langle\vec{\mathbf{r}}(t)\,\cev{\mathbf{r}}(t)\rangle = 2V + \Gm/(4\Gmeas).
\end{eqnarray}
This can be expressed simply as 
\begin{eqnarray}
   \sigma^2(t) = V + V_E,
\end{eqnarray}
the main result of this calculation. Very much compatible with intuition, statistical averaging over many experimental runs yields a classical variance given by the sum of the quantum uncertainty of the prediction, plus the quantum uncertainty of the retrodiction.

Let's now suppose that the prediction conditioning is stopped at time $t^*<t$ and after that the state $\rho(t>t^*)$ is propagated deterministically according to the master equation (as in the right half of Fig.~\ref{f:fig_4}). The trajectory becomes
\begin{eqnarray}\label{e:rt_sol_prop}
\vec{\mathbf{r}}(t>t^*) =  \vec{\mathbf{r}}(t^*)\times e^{-\frac{\Gm}{2}\left(t-t^*\right)}= \sqrt{4 \Gmeas}V \int_{-\infty}^{t^*} e^{-\frac{\Gm}{2}\left(t-\tau\right)} d\mathbf{W}(\tau).
\end{eqnarray}
This will change 
\begin{eqnarray}
\langle\vec{\mathbf{r}}(t)^2\rangle = 4\frac{\Gmeas}{\Gm} V^2 e^{-\Gm\left(t-t^*\right)}
\end{eqnarray}
and
\begin{eqnarray}
\langle\vec{\mathbf{r}}(t)\,\cev{\mathbf{r}}(t)\rangle = 4\frac{\Gmeas}{\Gm}V^2 e^{-\Gm\left(t-t^*\right)}.
\end{eqnarray}
The experimental variance becomes
\begin{eqnarray}
\sigma^2(t>t^*) = \frac{4\Gmeas}{\Gm} \left(V_E^2 -  V^2e^{-\Gm(t-t^*)}\right) = V + V_E + \frac{4\Gmeas}{\Gm} V^2 \left(1-e^{-\Gm(t-t^*)}\right).
\end{eqnarray}
That is, from the moment the measurement is ignored ($t^*$), the experimental variance rises exponentially from the minimum value $V+V_E$ to the much larger value $\sigma^2(t\gg t^*)\approx(4\Gmeas/\Gm) V^2\approx \bnth +1/2+ \Gqba/\Gm$, which coincides with the expected unconditional variance (assuming $\Gmeas\gg\Gm)$.

\paragraph*{Effect of demodulation filter}
When demodulating the homodyne photocurrent to obtain our quadrature measurement channels, $i_X$ and $i_Y$, a low-pass filter is applied.  It is important that this filter bandwidth is large enough that it does not attenuate the mechanical information contained in the signal.  (Or equivalently, that the filter time constant is much shorter than the mechanical dynamics of interest.)  The full bandwidth of this filter is 120kHz, and its effect on the homodyne spectrum is shown in Fig.~S1.

\begin{figure}
\begin{center}
\includegraphics[scale=1]{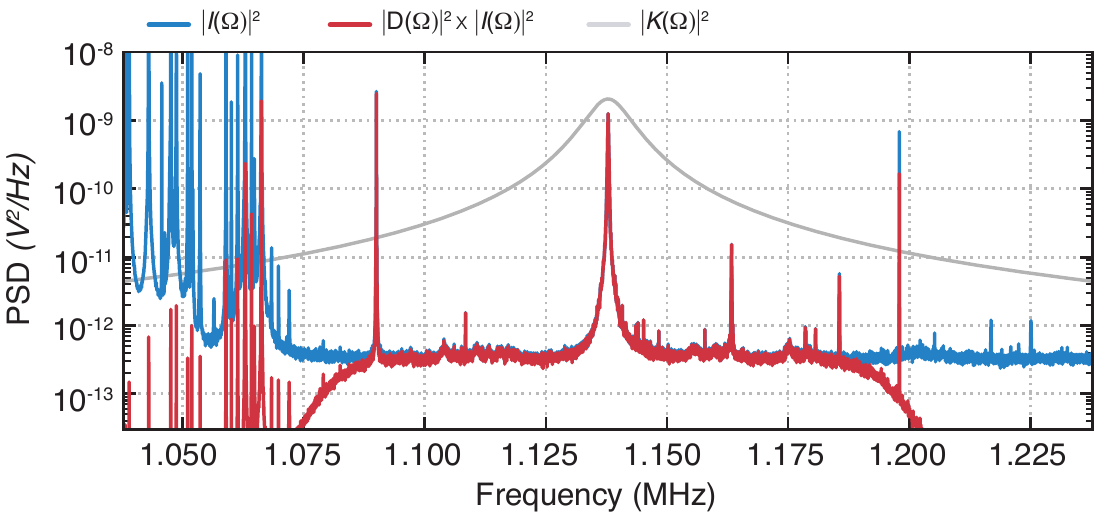}
\caption{{\bf Effect of the demodulation filter.}
Ensemble average power spectral density (PSD, blue) of the homodyne photocurrents $I$. The same photocurrents are demodulated at mechanical resonance frequency $\Om$. After demodulation, we apply a $2\times7^{th}$ order low-pass filter with a bandwidth of 60~kHz (equivalent to a 120~kHz bandpass filter). To illustrate that such filter does not affect the mechanical mode, we calculate the PSD of $i_X\cos{\Om t}+i_y\sin{\Om t}$ (red).  In grey we also show the steady-state filter for estimating the trajectory.
\label{f:methods_fig_1}}
\end{center}
\end{figure}

We can assess the effect of this demodulation filter on various statistical quantities, by rewriting $\vec{\mathbf{r}}(t)$ and $\cev{\mathbf{r}}(t)$ explicitly in terms of the measurement record, again assuming that the steady-state has been reached:
\begin{align*}
\vec{\mathbf{r}}(t) &= \sqrt{4\Gmeas}V\int_{-\infty}^te^{-\alpha (t-t')}\mathbf{i}(t')dt'
        \qquad&\qquad       \cev{\mathbf{r}}(t) &= -\sqrt{4\Gmeas}V_E\int_{\infty}^te^{\alpha(t-t')}\mathbf{i}(t')dt'\\
&= \sqrt{4\Gmeas}V\int_{0}^{\infty}e^{-\alpha \tau}\mathbf{i}(t-\tau)d\tau
        \quad& &= \sqrt{4\Gmeas}V_E\int_{-\infty}^{0}e^{\alpha \tau}\mathbf{i}(t-\tau)d\tau\\
&= \int_{-\infty}^{\infty}K(\tau)\mathbf{i}(t-\tau)d\tau
        \quad& &= \int_{-\infty}^{\infty}\overline{K}(\tau)\mathbf{i}(t-\tau)d\tau\\
&= (K*\mathbf{i})(t)
        \quad&\quad &= (\overline{K}*\mathbf{i})(t)
\end{align*}
where $K(t) = \sqrt{4\Gmeas}VH(t)e^{-\alpha t}$, $\overline{K}(t) = \sqrt{4\Gmeas}V_E H(-t)e^{\alpha t}$, $H(t)$ is the Heaviside step function, $\alpha=\Gm/2+4\Gmeas V$, and $*$ indicates convolution. Thus, in the steady state, we have both trajectories written as convolutions of the measurement record with a (Kalman)\cite{Doherty:1999ab} filter kernel, which is carefully defined to ensure the appropriate causality.

The measurement record itself can similarly be expressed as a convolution $\mathbf{i}(t) = (\mathbf{D}*I)(t)$ of the photocurrent, $I(t)$, with a bandpass filter kernel, $\mathrm{\mathbf{D}}$. Thus, we can easily express the trajectories as compound convolutions, or products of convolutions in the Fourier domain:
\begin{align*}
\vec{\mathbf{r}}(t) &= (K*\mathbf{D}*I)(t)          &   \cev{\mathbf{r}}(t) &= (\overline{K}*\mathbf{D}*I)(t)\\
\vec{\mathbf{r}}[\Og] &= K[\Og]\mathbf{D}[\Og]I[\Og]    &   \cev{\mathbf{r}}[\Og] &= \overline{K}[\Og]\mathbf{D}[\Og]I[\Og]\\
\end{align*}
The latter form is useful, since we can express various variances and covariances as integrals of the power spectra and cross-spectra, making use of the cross-correlation theorem for real, ergodic signals ($\left<f(t)g(t+\tau)\right> = \int_{-\infty}^{\infty}e^{i\Og \tau}\left<g[-\Og]f[\Og]\right>$ d\Og), evaluated for $\tau=0$.
\begin{align*}
    \left<\,\vec{\mathbf{r}}(t)\vec{\mathbf{r}}(t)\,\right> 
        &= \int_{-\infty}^{\infty}\left<|K[\Og]\mathbf{D}[\Og]I[\Og]|^2\right> d\Og\\
    \left<\,\cev{\mathbf{r}}(t)\cev{\mathbf{r}}(t)\,\right> 
        &= \int_{-\infty}^{\infty}\left<|\overline{K}[\Og]\mathbf{D}[\Og]I[\Og]|^2\right> d\Og\\
    \left<\,\vec{\mathbf{r}}(t)\cev{\mathbf{r}}(t)\,\right> 
        &= \int_{-\infty}^{\infty}\left<K[-\Og]\overline{K}[\Og]\,\bigl|\mathbf{D}[\Og]I[\Og]\bigr|^2\right> d\Og
\end{align*}

We can now easily compare these statistical quantities in the presence or absence of the filter, $\mathbf{D}$, using the above definitions of $K$, $\overline{K}$ and the known average power spectrum of $I[\Og]$.  The results are summarized in Table~S1.  Note that the difference in $\sigma^2$ reflects a systematic underestimate of the relative variance, which we correct for when presenting quantities (e.g. purity) in the main text.
\begin{table}
\centering
\begin{tabular}{c|c|c|c}
                                                            &   With Filter, $\mathbf{D}$       &   Without Filter          &       Difference\\
    \hline
    $\left<\,\vec{\mathbf{r}}(t)\,\vec{\mathbf{r}}(t)\,\right>$    &          21.20                        &         21.21       &          .05\%\\ 
    $\left<\,\cev{\mathbf{r}}(t)\,\cev{\mathbf{r}}(t)\,\right>$     &          22.42                        &          22.44       &          .09\%\\
    $\left<\,\vec{\mathbf{r}}(t)\,\cev{\mathbf{r}}(t)\,\right>$    &          21.23                        &          21.21       &          .09\%\\
    $\sigma^2=\left<\,(\vec{\mathbf{r}}(t)-\cev{\mathbf{r}}(t))^2\,\right>\quad$     &          1.17                        &          1.24       &          5.6\%\\

\end{tabular}
\caption{Variance and covariance in the presence and absence of demodulation filter $\mathbf{D}$.}
\label{tab:bogus_corr}
\end{table}

\paragraph*{System parameters and calibration\\}
\begin{table}[h]
\centering
\begin{tabular}{llll}
    \hline
    Symbol & Definition & Name & Value  \\ \hline
    $\Om$ & & Mechanical resonance frequency & $2\pi\times$1.138~MHz \\
    $Q_0$ & $\Om/\Gamma_0$ & Intrinsic mechanical quality factor & $1.03\times10^9$ \\
    $T$ & & Intrinsic mechanical bath temperature & $11\,\mathrm{K}$ \\
    $Q$ & $\Om/\Gm$ & Effective mechanical quality factor & $8740$ \\
    $\Gm$ &  & Effective mechanical damping rate & $2\pi\times$130~Hz \\
    $\meff$ & & Effective mass & 2.3~ng \\
    $\xzpf$ & $\sqrt{\frac{\hbar}{2\meff\Om}}$ & Displacement zero point fluctuations & 1.8~fm \\
    $\vcr$ & & Vacuum optomechanical coupling, probe & $2\pi\times129$~Hz\\
    $\kpr$ & & Linewidth of probe cavity mode & $2\pi\times$18.5~MHz\\
    $\Dpr$ & & Detuning of probe laser & $- 0.05\,\kpr<\Dpr<0$\\
    $\bncavpr$ & & Intracavity photons, probe laser & \\
    $\etac$ & & Cavity outcoupling, probe laser & $0.93$ \\
    $\eta_\mathrm{det}$ & & All-in homodyne detection efficiency & 0.74\\    
    $\Gamma_\mathrm{meas}$ & $4\eta_\mathrm{det}g^2/\kappa$ & Measurement rate & $2\pi\times$1.88~kHz\\
    $\Gamma_\mathrm{qba}$ & $4g^2/\kappa$ & Measurement-induced quantum backaction& $2\pi\times$2.54~kHz\\
    $\gamma$ & $\Gm \bnth $ & Thermal decoherence rate & $2\pi\times$0.26~kHz\\
    $\etameas$ & $\Gmeas/(\Gamma_\mathrm{qba}+\gamma)$ & Measurement efficiency & 68\% \\\hline
\end{tabular}
\caption{Parameters and definitions.}
\label{tab:parameters}
\end{table}
The optomechanical model underlying our stochastic master equation is based on relatively few parameters, most of which area easily obtained by independent measurements.  Cavity linewidths are determined by measuring the cavity response to a swept laser with calibrated sidebands.  The mechanical quality factor is determined by exciting coherent motion and monitoring the exponential ringdown.  This ringdown measurement is made using a laser which addresses a very-low-finesse cavity mode, such that it has negligible optomechanical coupling.  By repeating this ringdown measurement under various conditions, including making ``stroboscopic'' ringdowns, we can confidently rule out any optomechanical dynamical backaction in this $Q$ measurement.  Additional details on this $Q$ measurement can be found in Ref. [\citen{Rossi:2018ab}].  

Our optomechanical coupling rate, $g_0$, is determined by a series of optomechanically induced transparency (OMIT) measurements.  This allows for a coherent response measurement of the mechanical oscillator, from which the coupling rate can be inferred.  In prior works (using the exact same experimental system), we have also made $g_0$ calibration measurements using an independent technique based on quantum backaction, and found good agreement with the OMIT method, lending support to this calibration technique.  Additional details on both calibration methods can be found in Ref. [\citen{Rossi:2018ab}]. From the background of the average PSD calibrated in absolute displacement units (through phase modulation technique) and using standard optomechanical theory we estimate a total detection efficiency $\etadet=0.74$, consistent with independent measurements of the optical losses in the detection path. 
In addition, the level of this background is also used to normalize the photocurrent such that the PSD has a background of 1. This calibration is needed when one wants to use the photocurrent to calculate the quantum trajectories and POVM measurement according to eqs.~\eqref{e:Z_dyn} and \eqref{e:r_E}.


\clearpage
\newpage

\begin{thebibliography}{10}
\expandafter\ifx\csname url\endcsname\relax
  \def\url#1{\texttt{#1}}\fi
\expandafter\ifx\csname urlprefix\endcsname\relax\def\urlprefix{URL }\fi
\providecommand{\bibinfo}[2]{#2}
\providecommand{\eprint}[2][]{\url{#2}}

\bibitem{Braginsky:1992aa}
\bibinfo{author}{Braginsky, V.~B.} \& \bibinfo{author}{Khalili, F.~Y.}
\newblock \emph{\bibinfo{title}{Quantum Measurement}}
  (\bibinfo{publisher}{Cambridge University Press}, \bibinfo{year}{1992}).

\bibitem{Wiseman:2010aa}
\bibinfo{author}{Wiseman, H.~M.} \& \bibinfo{author}{Milburn, G.~J.}
\newblock \emph{\bibinfo{title}{Quantum Measurement and Control}}
  (\bibinfo{publisher}{Cambridge University Press}, \bibinfo{year}{2010}).

\bibitem{Guerlin:2007aa}
\bibinfo{author}{Guerlin, C.} \emph{et~al.}
\newblock \bibinfo{title}{Progressive field-state collapse and quantum
  non-demolition photon counting}.
\newblock \emph{\bibinfo{journal}{Nature}} \textbf{\bibinfo{volume}{448}},
  \bibinfo{pages}{889--894} (\bibinfo{year}{2007}).

\bibitem{Murch:2013aa}
\bibinfo{author}{Murch, K.~W.}, \bibinfo{author}{Weber, S.~J.},
  \bibinfo{author}{Macklin, C.} \& \bibinfo{author}{Siddiqi, I.}
\newblock \bibinfo{title}{Observing single quantum trajectories of a
  superconducting quantum bit}.
\newblock \emph{\bibinfo{journal}{Nature}} \textbf{\bibinfo{volume}{502}},
  \bibinfo{pages}{211--214} (\bibinfo{year}{2013}).

\bibitem{Weber:2014aa}
\bibinfo{author}{Weber, S.~J.} \emph{et~al.}
\newblock \bibinfo{title}{Mapping the optimal route between two quantum
  states}.
\newblock \emph{\bibinfo{journal}{Nature}} \textbf{\bibinfo{volume}{511}},
  \bibinfo{pages}{570--573} (\bibinfo{year}{2014}).

\bibitem{Doherty:2012aa}
\bibinfo{author}{Doherty, A.~C.}, \bibinfo{author}{Szorkovszky, A.},
  \bibinfo{author}{Harris, G.~I.} \& \bibinfo{author}{Bowen, W.~P.}
\newblock \bibinfo{title}{The quantum trajectory approach to quantum feedback
  control of an oscillator revisited}.
\newblock \emph{\bibinfo{journal}{Phil. Tran. R. Soc. A}}
  \textbf{\bibinfo{volume}{370}}, \bibinfo{pages}{5338--5353}
  (\bibinfo{year}{2012}).

\bibitem{Doherty:1999ab}
\bibinfo{author}{Doherty, A.~C.}, \bibinfo{author}{Tan, S.~M.},
  \bibinfo{author}{Parkins, A.~S.} \& \bibinfo{author}{Walls, D.~F.}
\newblock \bibinfo{title}{State determination in continuous measurement}.
\newblock \emph{\bibinfo{journal}{Phys. Rev. A}} \textbf{\bibinfo{volume}{60}},
  \bibinfo{pages}{2380--2392} (\bibinfo{year}{1999}).

\bibitem{Bowen:2016aa}
\bibinfo{author}{Bowen, W.~P.} \& \bibinfo{author}{Milburn, G.~J.}
\newblock \emph{\bibinfo{title}{Quantum Optomechanics}}
  (\bibinfo{publisher}{CRC Press}, \bibinfo{year}{2016}).

\bibitem{Lammers:2018aa}
\bibinfo{author}{Lammers, J.}
\newblock \emph{\bibinfo{title}{State preparation and verification in
  continuously measured quantum systems}}.
\newblock Ph.D. thesis, \bibinfo{school}{Leibniz Universit{\"a}t Hannover}
  (\bibinfo{year}{2018}).

\bibitem{Lammers:2018ab}
\bibinfo{author}{Lammers, J.} \& \bibinfo{author}{Hammerer, K.}
\newblock \bibinfo{title}{In preparation} .

\bibitem{Jacobs:2006aa}
\bibinfo{author}{Jacobs, K.} \& \bibinfo{author}{Steck, D.~A.}
\newblock \bibinfo{title}{A straightforward introduction to continuous quantum
  measurement}.
\newblock \emph{\bibinfo{journal}{Contemp. Phys.}}
  \textbf{\bibinfo{volume}{47}}, \bibinfo{pages}{279--303}
  (\bibinfo{year}{2006}).

\bibitem{Aspelmeyer:2014aa}
\bibinfo{author}{Aspelmeyer, M.}, \bibinfo{author}{Kippenberg, T.~J.} \&
  \bibinfo{author}{Marquardt, F.}
\newblock \bibinfo{title}{Cavity optomechanics}.
\newblock \emph{\bibinfo{journal}{Rev. Mod. Phys.}}
  \textbf{\bibinfo{volume}{86}}, \bibinfo{pages}{1392} (\bibinfo{year}{2014}).

\bibitem{Clerk:2010ab}
\bibinfo{author}{Clerk, A.~A.}, \bibinfo{author}{Devoret, M.~H.},
  \bibinfo{author}{Girvin, S.~M.}, \bibinfo{author}{Marquardt, F.} \&
  \bibinfo{author}{Schoelkopf, R.~J.}
\newblock \bibinfo{title}{Introduction to quantum noise, measurement, and
  amplification}.
\newblock \emph{\bibinfo{journal}{Rev. Mod. Phys.}}
  \textbf{\bibinfo{volume}{82}}, \bibinfo{pages}{1155--1208}
  (\bibinfo{year}{2010}).

\bibitem{Vanner:2013aa}
\bibinfo{author}{Vanner, M.~R.}, \bibinfo{author}{Hofer, J.},
  \bibinfo{author}{Cole, G.~D.} \& \bibinfo{author}{Aspelmeyer, M.}
\newblock \bibinfo{title}{Cooling-by-measurement and mechanical state
  tomography via pulsed optomechanics}.
\newblock \emph{\bibinfo{journal}{Nature Communications}}
  \textbf{\bibinfo{volume}{4}}, \bibinfo{pages}{2295} (\bibinfo{year}{2013}).

\bibitem{Wieczorek:2015aa}
\bibinfo{author}{Wieczorek, W.} \emph{et~al.}
\newblock \bibinfo{title}{Optimal state estimation for cavity optomechanical
  systems}.
\newblock \emph{\bibinfo{journal}{Phys. Rev. Lett.}}
  \textbf{\bibinfo{volume}{114}}, \bibinfo{pages}{223601}
  (\bibinfo{year}{2015}).

\bibitem{Harris:2016aa}
\bibinfo{author}{Harris, G.~I.} \emph{et~al.}
\newblock \bibinfo{title}{Laser cooling and control of excitations in
  superfluid helium}.
\newblock \emph{\bibinfo{journal}{Nature Physics}}
  \textbf{\bibinfo{volume}{12}}, \bibinfo{pages}{788 EP --}
  (\bibinfo{year}{2016}).

\bibitem{Setter:2018aa}
\bibinfo{author}{Setter, A.}, \bibinfo{author}{Toro\ifmmode~\check{s}\else
  \v{s}\fi{}, M.}, \bibinfo{author}{Ralph, J.~F.} \& \bibinfo{author}{Ulbricht,
  H.}
\newblock \bibinfo{title}{Real-time kalman filter: Cooling of an optically
  levitated nanoparticle}.
\newblock \emph{\bibinfo{journal}{Phys. Rev. A}} \textbf{\bibinfo{volume}{97}},
  \bibinfo{pages}{033822} (\bibinfo{year}{2018}).

\bibitem{Tan:2015aa}
\bibinfo{author}{Tan, D.}, \bibinfo{author}{Weber, S.~J.},
  \bibinfo{author}{Siddiqi, I.}, \bibinfo{author}{M\o{}lmer, K.} \&
  \bibinfo{author}{Murch, K.~W.}
\newblock \bibinfo{title}{Prediction and retrodiction for a continuously
  monitored superconducting qubit}.
\newblock \emph{\bibinfo{journal}{Phys. Rev. Lett.}}
  \textbf{\bibinfo{volume}{114}}, \bibinfo{pages}{090403}
  (\bibinfo{year}{2015}).

\bibitem{Zhang:2017aa}
\bibinfo{author}{Zhang, J.} \& \bibinfo{author}{M\o{}lmer, K.}
\newblock \bibinfo{title}{Prediction and retrodiction with continuously
  monitored gaussian states}.
\newblock \emph{\bibinfo{journal}{Phys. Rev. A}} \textbf{\bibinfo{volume}{96}},
  \bibinfo{pages}{062131} (\bibinfo{year}{2017}).

\bibitem{Gammelmark:2013aa}
\bibinfo{author}{Gammelmark, S.}, \bibinfo{author}{Julsgaard, B.} \&
  \bibinfo{author}{M\o{}lmer, K.}
\newblock \bibinfo{title}{Past quantum states of a monitored system}.
\newblock \emph{\bibinfo{journal}{Phys. Rev. Lett.}}
  \textbf{\bibinfo{volume}{111}}, \bibinfo{pages}{160401}
  (\bibinfo{year}{2013}).

\bibitem{Rybarczyk:2015aa}
\bibinfo{author}{Rybarczyk, T.} \emph{et~al.}
\newblock \bibinfo{title}{Forward-backward analysis of the photon-number
  evolution in a cavity}.
\newblock \emph{\bibinfo{journal}{Phys. Rev. A}} \textbf{\bibinfo{volume}{91}},
  \bibinfo{pages}{062116} (\bibinfo{year}{2015}).

\bibitem{Tsaturyan:2017aa}
\bibinfo{author}{Tsaturyan, Y.}, \bibinfo{author}{Barg, A.},
  \bibinfo{author}{Polzik, E.~S.} \& \bibinfo{author}{Schliesser, A.}
\newblock \bibinfo{title}{Ultracoherent nanomechanical resonators via soft
  clamping and dissipation dilution}.
\newblock \emph{\bibinfo{journal}{Nat Nano}} \textbf{\bibinfo{volume}{12}},
  \bibinfo{pages}{776--783} (\bibinfo{year}{2017}).

\bibitem{Rossi:2018ab}
\bibinfo{author}{Rossi, M.}, \bibinfo{author}{Mason, D.},
  \bibinfo{author}{Chen, J.}, \bibinfo{author}{Tsaturyan, Y.} \&
  \bibinfo{author}{Schliesser, A.}
\newblock \bibinfo{title}{Measurement-based quantum control of mechanical
  motion}.
\newblock \emph{\bibinfo{journal}{Nature}} \textbf{\bibinfo{volume}{563}},
  \bibinfo{pages}{53--58} (\bibinfo{year}{2018}).

\bibitem{Belavkin:1980aa}
\bibinfo{author}{Belavkin, V.~P.}
\newblock \bibinfo{title}{Optimal filtering of markov signals with quantum
  white noise}.
\newblock \emph{\bibinfo{journal}{Radio Eng. Electron. Phys. (USSR)}}
  \textbf{\bibinfo{volume}{25}} (\bibinfo{year}{1980}).

\bibitem{Hofer:2015ab}
\bibinfo{author}{Hofer, S.~G.} \& \bibinfo{author}{Hammerer, K.}
\newblock \bibinfo{title}{Entanglement-enhanced time-continuous quantum control
  in optomechanics}.
\newblock \emph{\bibinfo{journal}{Phys. Rev. A}} \textbf{\bibinfo{volume}{91}},
  \bibinfo{pages}{033822} (\bibinfo{year}{2015}).

\bibitem{Doherty:1999aa}
\bibinfo{author}{Doherty, A.~C.} \& \bibinfo{author}{Jacobs, K.}
\newblock \bibinfo{title}{Feedback control of quantum systems using continuous
  state estimation}.
\newblock \emph{\bibinfo{journal}{Phys. Rev. A}} \textbf{\bibinfo{volume}{60}},
  \bibinfo{pages}{2700--2711} (\bibinfo{year}{1999}).

\bibitem{Danilishin:2013aa}
\bibinfo{author}{Danilishin, S.}, \bibinfo{author}{Miao, H.},
  \bibinfo{author}{M{\"u}ller-Ebhardt, H.} \& \bibinfo{author}{Chen, Y.}
\newblock \bibinfo{title}{Optomechanical entanglement: How to prepare, verify
  and "steer" a macroscopic mechanical quantum state?}
\newblock In \emph{\bibinfo{booktitle}{Proceedings of the First International
  Workshop on ECS and Its Application to QIS}} (\bibinfo{year}{2013}).

\bibitem{Nimmrichter:2014aa}
\bibinfo{author}{Nimmrichter, S.}, \bibinfo{author}{Hornberger, K.} \&
  \bibinfo{author}{Hammerer, K.}
\newblock \bibinfo{title}{Optomechanical sensing of spontaneous wave-function
  collapse}.
\newblock \emph{\bibinfo{journal}{Phys. Rev. Lett.}}
  \textbf{\bibinfo{volume}{113}}, \bibinfo{pages}{020405}
  (\bibinfo{year}{2014}).

\bibitem{Bassi:2017aa}
\bibinfo{author}{Bassi, A.}, \bibinfo{author}{Gro{\ss}ardt, A.} \&
  \bibinfo{author}{Ulbricht, H.}
\newblock \bibinfo{title}{Gravitational decoherence}.
\newblock \emph{\bibinfo{journal}{Class. Quantum Grav.}}
  \textbf{\bibinfo{volume}{34}}, \bibinfo{pages}{193002}
  (\bibinfo{year}{2017}).

\bibitem{McMillen:2017aa}
\bibinfo{author}{McMillen, S.} \emph{et~al.}
\newblock \bibinfo{title}{Quantum-limited estimation of continuous spontaneous
  localization}.
\newblock \emph{\bibinfo{journal}{Phys. Rev. A}} \textbf{\bibinfo{volume}{95}},
  \bibinfo{pages}{012132} (\bibinfo{year}{2017}).

\end{thebibliography}
\end{document}